\documentclass[usegraphicx]{mn2e}

\usepackage{times}
\usepackage{amssymb}
\usepackage{amsmath}
\begin{document}

\include{defn}
\def\cm{{\rm\thinspace cm}}
\def\dyn{{\rm\thinspace dyn}}
\def\erg{{\rm\thinspace erg}}
\def\eV{{\rm\thinspace eV}}
\def\MeV{{\rm\thinspace MeV}}
\def\g{{\rm\thinspace g}}
\def\ga{{\rm\thinspace gauss}}
\def\K{{\rm\thinspace K}}
\def\keV{{\rm\thinspace keV}}
\def\km{{\rm\thinspace km}}
\def\kpc{{\rm\thinspace kpc}}
\def\Lsun{\hbox{$\rm\thinspace L_{\odot}$}}
\def\m{{\rm\thinspace m}}
\def\Mpc{{\rm\thinspace Mpc}}
\def\Msun{\hbox{$\rm\thinspace M_{\odot}$}}
\def\pc{{\rm\thinspace pc}}
\def\ph{{\rm\thinspace ph}}
\def\s{{\rm\thinspace s}}
\def\yr{{\rm\thinspace yr}}
\def\sr{{\rm\thinspace sr}}
\def\Hz{{\rm\thinspace Hz}}
\def\MHz{{\rm\thinspace MHz}}
\def\GHz{{\rm\thinspace GHz}}
\def\chisq{\hbox{$\chi^2$}}
\def\delchi{\hbox{$\Delta\chi$}}
\def\cmps{\hbox{$\cm\s^{-1}\,$}}
\def\cmpssq{\hbox{$\cm\s^{-2}\,$}}
\def\cmsq{\hbox{$\cm^2\,$}}
\def\cmcu{\hbox{$\cm^3\,$}}
\def\pcmcu{\hbox{$\cm^{-3}\,$}}
\def\pcmcuK{\hbox{$\cm^{-3}\K\,$}}
\def\dynpcmsq{\hbox{$\dyn\cm^{-2}\,$}}
\def\ergcmcups{\hbox{$\erg\cm^3\ps\,$}}
\def\ergpcmps{\hbox{$\erg\cm^{-3}\s^{-1}\,$}}
\def\ergpcmsqps{\hbox{$\erg\cm^{-2}\s^{-1}\,$}}
\def\ergpcmsqpspA{\hbox{$\erg\cm^{-2}\s^{-1}$\AA$^{-1}\,$}}
\def\ergpcmsqpspsr{\hbox{$\erg\cm^{-2}\s^{-1}\sr^{-1}\,$}}
\def\ergpcmcups{\hbox{$\erg\cm^{-3}\s^{-1}\,$}}
\def\ergpcmps{\hbox{$\erg\cm^{-1}\s^{-1}$}}
\def\ergps{\hbox{$\erg\s^{-1}\,$}}
\def\ergpspmp{\hbox{$\erg\s^{-1}\Mpc^{-3}\,$}}
\def\gpcm{\hbox{$\g\cm^{-3}\,$}}
\def\gpcmps{\hbox{$\g\cm^{-3}\s^{-1}\,$}}
\def\gps{\hbox{$\g\s^{-1}\,$}}
\def\Jy{{\rm Jy}}
\def\keVpcmsqpspsr{\hbox{$\keV\cm^{-2}\s^{-1}\sr^{-1}\,$}}
\def\kmps{\hbox{$\km\s^{-1}\,$}}
\def\kmpspmp{\hbox{$\km\s^{-1}\Mpc{-1}\,$}}
\def\Lsunppc{\hbox{$\Lsun\pc^{-3}\,$}}
\def\Msunpc{\hbox{$\Msun\pc^{-3}\,$}}
\def\Msunpkpc{\hbox{$\Msun\kpc^{-1}\,$}}
\def\Msunppc{\hbox{$\Msun\pc^{-3}\,$}}
\def\Msunppcpyr{\hbox{$\Msun\pc^{-3}\yr^{-1}\,$}}
\def\Msunpyr{\hbox{$\Msun\yr^{-1}\,$}}
\def\pcm{\hbox{$\cm^{-3}\,$}}
\def\pcmsq{\hbox{$\cm^{-2}\,$}}
\def\pcmK{\hbox{$\cm^{-3}\K$}}
\def\phpcmsqps{\hbox{$\ph\cm^{-2}\s^{-1}\,$}}
\def\pHz{\hbox{$\Hz^{-1}\,$}}
\def\pmpc{\hbox{$\Mpc^{-1}\,$}}
\def\pmpccu{\hbox{$\Mpc^{-3}\,$}}
\def\ps{\hbox{$\s^{-1}\,$}}
\def\psqcm{\hbox{$\cm^{-2}\,$}}
\def\psr{\hbox{$\sr^{-1}\,$}}
\def\kmpspMpc{\hbox{$\kmps\Mpc^{-1}$}}

\title{On conduction, cooling flows and galaxy formation}
\author[]
{\parbox[]{6.in} {A.C. Fabian, L.M. Voigt and R.G. Morris\\
\footnotesize
Institute of Astronomy, Madingley Road, Cambridge CB3 0HA\\
}}

\maketitle
\begin{abstract}
On the basis of the universal gas fraction in clusters of galaxies, we
estimate that the effective thermal conductivity required to balance
radiative cooling in the cores, where the gas temperature is
3--10~keV, is about one tenth of the Spitzer rate. This confirms that
thermal conduction can be important for the energy balance provided
that it is not highly suppressed by magnetic fields in the gas. We
determine the global effective conductivity in a sample of 29 clusters
using published X-ray data on the inferred cooling rates and show that
most lie between one and one tenth of the Spitzer rate. More work on
the profiles in cooling flow clusters is required to test the
conduction hypothesis further. We examine the possibility that
conduction operates during galaxy formation, and show that it provides
a simple explanation for the upper-mass cutoff in galaxy masses.
\end{abstract}
\begin{keywords}
galaxies: clusters -- cooling flows -- X-rays: galaxies -- conduction
\end{keywords}

\section{Introduction}

Recent X-ray data on the cores of clusters of galaxies from
\textit{XMM-Newton} and \textit{Chandra} indicate that some heat flux must
be balancing the radiative cooling losses of the gas (Peterson et al.\
2001; Tamura et al.\ 2001; Fabian et al.\ 2001; Johnstone et al.\ 2002,
B\"ohringer et al.\ 2002; McNamara et al.\ 2001; Nulsen et al.\ 2002). The
source of heat is unclear, with the central radio source being one
possibility (Churazov et al.\ 2002; Br\"uggen \& Kaiser 2001; Reynolds et
al.\ 2002) and conduction being another (Narayan \& Medvedev 2001; Gruzinov
2002; Voigt et al.\ 2002). Here we examine conduction and its consequences
in more detail.

The observations show that large temperature drops, by a factor of
three or more, are common close to the centre of clusters. Thermal
conduction is therefore an obvious heat transport process. The
question is one of the level of conductivity and whether it can be
high enough to balance radiative losses, and also whether it can do so
in a stable manner over the long time-scales required. For a highly
ionized plasma such as the intracluster medium the maximum rate is
expected to be that calculated by Spitzer (1962), which we hereafter
refer to as Spitzer conductivity with coefficient $\kappa_{\rm S}$.
Magnetic fields in the plasma should suppress the conductivity by some
factor $f$ so that $\kappa=f\kappa_{\rm S}$. The value of $f$ is not
clear, but could be as high as one third for a tangled field (see
Narayan \& Medvedev 2001 and Malyshkin 2001 and references therein).
Suppression should be very high across field lines and appears to be
high across the very abrupt temperature drops known as cold fronts
found in some clusters (Ettori \& Fabian 2000; Markevich et al.\ 2000;
Vikhlinin et al.\ 2001).

In Section~2 we show that the level of conductivity required to
balance radiative cooling in the core of a cluster, where the
radiative cooling time is less than the age of the Universe, is less
than $\kappa_{\rm S}$. This means that if conduction is unimpeded then
it can balance radiation. We then compare this result with an effective
conductivity from observations, determined by
\begin{equation}
\kappa_{\rm eff}= {{L(<r)}\over{4\pi r^2 dT/dr}},
\end{equation}
where $L$ is the total luminosity within radius $r$ at which the
temperature gradient $dT/dr$ is inferred. The sample used is from the
brightest 55 in the sky studied using \textit{ROSAT} data by Peres et
al.\ (1998), supplemented by some luminous clusters from the sample of
Allen (2000). We choose objects with known central emission-line
nebulosities, indicating probable large temperature drops in the gas.

If thermal conduction (i.e.\ electron motion) is relatively unimpeded
in a radial direction, this forces us to consider that ion motion
might not be as restricted as normally assumed. In other words,
sedimentation of heavy ions is a possibility, with implications which
we consider in Section~3.

Finally we discuss the influence of conduction on galaxy formation,
showing that it cannot balance cooling in small-to-medium mass
galaxies but can have a significant effect on massive galaxies. It may
therefore be responsible for the high-mass cutoff to galaxy masses.

\section{The effective conductivity $\kappa_{\rm \lowercase{eff}}$}

Consider the gas in the core of a cluster. Within radius $r$ gas is
radiatively emitting energy at a rate $L$ leading to a decreasing
temperature unless heat is supplied at a similar rate, which for
conductivity means $L=4\pi r^2 \kappa {{dT}\over{dr}}$. Ignoring a
small factor which depends on the density gradients within that
radius, and approximating $dT/dr$ as $aT/r$ (i.e.\ $T\propto r^a$), the
effective conductivity required is
\begin{equation}
\kappa_{\rm eff}\approx {{n_e^2 \Lambda r^2}\over{3Ta}}.
\end{equation}
$\Lambda$ is the radiative cooling function and $n_{e}$ is the r.m.s.\
electron density within $r$. Now, the radiative cooling time is given
by
\begin{equation}\label{eqn:tcool}
t_{\rm cool}\approx {3\over2}{nkT/n_e^2\Lambda}
\end{equation}
{where n is the gas density}, and so
\begin{equation}
\kappa_{\rm eff}\approx {{nkr^2}\over{2at_{\rm cool}}}.
\end{equation} 
From the virial theorem, $GM/r\approx kT/m$, where $m$ is the mean molecular
weight, so for a gas mass fraction $f_{\rm gas}$ we have
\begin{equation}\label{eqn:kappatheo}
\kappa_{\rm eff}={{f_{\rm gas}k^2 T}\over{8m^2G a t_{\rm cool}}}.
\end{equation}
As we shall see, for typical cluster parameters this is less than, but
reasonably close to, $\kappa_{\rm S}$.

We next consider the effective conductivity required by data,
$\kappa_{\rm eff}'$, for a sample of cooling flow clusters drawn from
the study of Peres et al.\ (1998). We use those objects for which
optical emission lines have been detected from the central cluster
galaxy, indicating that some gas has cooled out there. Peres et al.\
(1998) tabulate $L(<r_{\rm cool})$, where $r_{\rm cool}$ is the radius
at which $t_{\rm cool}={2\over3}H_0^{-1}=13$~Gyr, as well as the outer
(wide-beam) temperature of each cluster. Assuming that $dT/dr=0.4
T/r$, which is appropriate for conduction balancing
bremsstrahlung cooling above $\sim$2keV in a spherical, constant
pressure cluster (Fabian et al.\ 1994), and is
also a reasonable fit to observed temperature gradients, we plot in
Fig.~1
\begin{equation}\label{eqn:kappaobs}
  \kappa_{\rm eff}'={{L(<r_{\rm cool})}\over{1.6\pi r_{\rm cool}^2
      T/r_{\rm cool}}}.
\end{equation}
The sample has been supplemented to a total of 29 by some massive
clusters from the work of Allen (2000), where the integrated mass
deposition rate within the cooling radius is again calculated from
imaging data (we use IRAS 09104+4109, Abell 963, Zwicky 3146, Abell
1068, RXJ1347.5, Abell 1835, MS2137.3-2353, Abell 2390 and MS
1455.0+223). All exhibit optical emission lines. Where the
measured temperature of the cluster is greater than 10keV, the value is
taken from Allen et al.\ (2001).

\begin{figure}
\includegraphics[angle=-90, width=0.95\columnwidth]{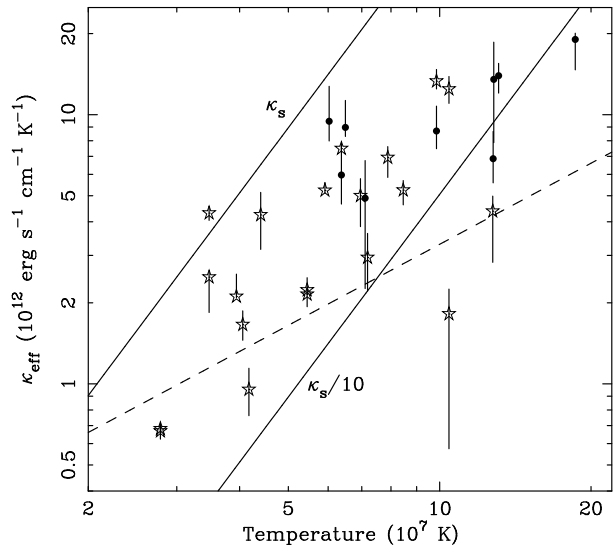}
\caption{$\kappa_{\rm eff}$ determined from Equation~\ref{eqn:kappaobs}
using data from (stars) Peres et al.\ (1998) for clusters which have
optical line emission noted for the central cluster galaxy, and from
(solid dots) Allen (2000). Lines for $\kappa_{\rm S}$ and one tenth
$\kappa_{\rm S}$ are shown, together with Equation~\ref{eqn:kappatheo}
(dashed), for the case $f_{\rm gas} = 0.17$, $a = 0.4$.}
\end{figure}

The bolometric luminosity emitted from within the cooling radius is
calculated from
\begin{equation}
L(<r_{\rm cool})= \frac{5kT}{m} \dot M
\end{equation}
where a factor of 2 is used to take into account the change in
gravitational potential energy of the inflowing mass which is assumed 
in the estimates of Allen (2000). 

Fig.~1 shows that most cooling flow clusters have $\kappa_{\rm eff}'$
between $\kappa_{\rm S}$ and $\kappa_{\rm S}/10$. Conduction may
therefore be important, as noted by Narayan \& Medvedev (2001) and
Gruzinov (2002). Our inclusion of the relation from
Equation~\ref{eqn:kappatheo}, however, shows that {\it all} clusters
where the central cooling time is less than a Hubble time and which
have the universal gas fraction must lie in this region too. It may,
therefore, be a coincidence that $\kappa_{\rm eff}'$ for clusters is
close to $\kappa_{\rm S}$.

Conversely, the results of Equation~\ref{eqn:kappatheo} indicate that
clusters lie in an interesting regime, where conduction is potentially very
effective, if it operates. Simple global estimates like those plotted in
Fig.~1, however, are poor diagnostics as to whether conduction is actually
taking place. Detailed study of the temperature profiles is required to
determine whether conduction is important or not. We have begun such tests
using \textit{Chandra} data on cluster cores (Voigt et al.\ 2002), which
supports conduction being an important factor in the energy flow from about
100~kpc into the innermost 20~kpc. Within the smaller radius, cooling may
dominate and lead to the observed optical nebulosities and star formation.

Loeb (2002) has noted that if conduction operates at the Spitzer rate
throughout clusters and beyond, they would leak their thermal energy
into the surrounding intergalactic medium. Therefore Spitzer rate
conduction may be restricted to a cluster core. Perhaps the formation
of a cooling flow and consequent gas infall leads to magnetic field
reconnection and reordering and to a more radial field (Soker \&
Sarazin 1990) which then allows conduction to become important (a
possibility mentioned by Bregman \& David 1989). In other words the
cooling flow produces the conditions which allow conduction to operate.

\section{Implications for sedimentation}

We note that if thermal conduction is uninhibited by magnetic fields
then so may be sedimentation within the cluster core. This could occur
if, for example, a significant fraction of the magnetic fields were
radial. Basically this means that ions heavier than protons will
accumulate towards the centre relative to hydrogen. Although iron may
therefore accumulate (Fabian \& Pringle 1977) the effect is more
important for helium nuclei (Gilfanov \& Sunyaev 1984). Calculations
have recently been performed for a Navarro, Frenk \& White (1997)
potential by Qin \& Wu (2000). The abundance gradients seen in iron,
and some other elements, in some cluster cores (Fukazawa et al.\ 1994;
Ezawa et al.\ 1997, De Grandi \& Molendi 2001; Sanders \& Fabian 2002;
Johnstone et al.\ 2002) may therefore be partly accounted for.

If helium does undergo sedimentation, then the inferred gas  density
in the central cooling regions is reduced. Any stars formed from cooled
gas will be helium rich and so have short main-sequence lifetimes
(Lynden-Bell 1967). Such a radical possibility would imply significant
increases in the star formation rate associated with the observed
distributed blue light (Johnstone, Fabian \& Nulsen 1987; Allen 1995;
Cardiel et al.\ 1998; Crawford et al.\ 1999). This will be studied further
in later work. 

We note that if conduction is suppressed then so will be
sedimentation, but if conduction is balancing cooling then
sedimentation does not necessarily operate, unless the field lines are
essentially radial.

\section{Galaxy formation}

Most models for galaxy formation are based on gas falling into dark
matter potential wells, being heated and then cooling radiatively to
form stars (e.g.\ White \& Frenk 1991). For massive galaxies the
cooling process resembles a cooling flow since the gas does not cool
immediately but forms a slowly cooling and settling atmosphere. A
problem arises then in that there is no simple upper mass limit to a
galaxy, and a supermassive galaxy could in principle form at the
centre of a cluster as a result of a massive cooling flow. In
practice, the models assume that visible stars do not form in potentials
above some mass (e.g.\ Kauffmann et al.\ 1999). We now investigate the
effect of conduction on this cooling process.

\begin{figure}
\includegraphics[angle=-90, width=0.95\columnwidth]{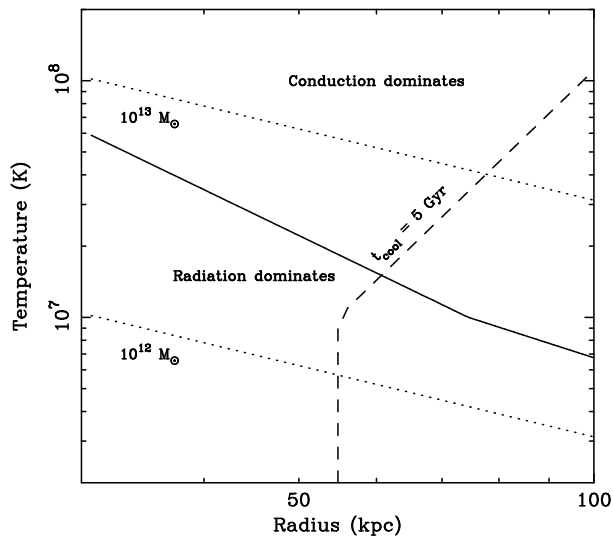}
\caption{Regime (above solid line) where conduction dominates and
(below) where radiative cooling dominates. An isothermal potential has been
assumed and lines of constant mass (from the virial theorem) are plotted
(dotted lines). It explains why gas cannot cool sufficiently to make
massive galaxies above a few times $10^{12}\Msun$. Along the dashed line
the radiative cooling time is 5~Gyr.}
\end{figure}

Following the models of White \& Frenk (1991) and Kauffmann et al.\
(1999) we adopt a simple isothermal distribution for the dark matter
and give the gas a similar density distribution;
\begin{equation}
n_{\rm e}\approx{{f_{\rm gas} M_0}\over{8\pi r_0 mr^2}},
\end{equation}
where $M_0$ is the total mass within radius $r_0$. The
luminosity within $r_0$ is then
\begin{equation}
L={{f_{\rm gas}^2M_0^2\Lambda}\over{32\pi r_0^3 m^2}}.
\end{equation}
We have assumed that gas extends only in to a radius of two thirds
$r_0$ in order that the luminosity is not dominated by an unphysical
central cusp. (We justify this factor from the expression for the
cooling rate in the models of Kauffmann et al.\ 1999 of $\dot M=4\pi n
m r_{\rm c}^2 dr_{\rm c}/dt$, $r_{\rm c}$ being the cooling radius at
time $t$, which with $L={3\over 2}{{\dot M}\over m}kT$ gives the above
value.)

Setting 
\begin{equation}
M_0={{kTr_0}\over{Gm}}
\end{equation} 
from the virial theorem, and balancing $L$ by a conductive flux gives 
\begin{equation}
{{f_{\rm gas}^2\Lambda}\over{32\pi m^2 r^3}}\left({{kTr}\over{Gm}}\right)^2=
4\pi r^2\kappa{{dT}\over{dr}}.
\end{equation} 
Let $dT/dr=aT/r,$ $\kappa=\kappa_0 T^{5/2}$ and for
bremsstrahlung $\Lambda=\Lambda_0 T^{1/2}$ then 
\begin{equation}
T={{f_{\rm gas}^2 \Lambda_0 k^2}\over{8(4\pi r)^2 \kappa_0 m^4 G^2 a}}.
\end{equation} 
This is approximately
\begin{equation}
T\approx 6\times 10^7 \left({r\over{\rm 30 kpc}}\right)^{-2}
\left({f\over{0.1}}\right)^2{\rm K}.
\end{equation} 
If cooling by metals is important ($T \la 10^7{\rm K}$) then
$\Lambda=\Lambda_{\rm m}$ and
\begin{equation}
T=\left({{f^2 \Lambda_{\rm m} k^2}\over{8(4\pi r_0)^2 \kappa_0 m^4 G^2
a}}\right)^{2/3}
\end{equation}
so
\begin{equation}
T\approx 3\times 10^7 \left({r\over{\rm 30 kpc}}\right)^{-4/3}
\left({f\over{0.1}}\right)^{4/3}{\rm K}.
\end{equation}

\begin{figure}
\includegraphics[angle=-90, width=0.95\columnwidth]{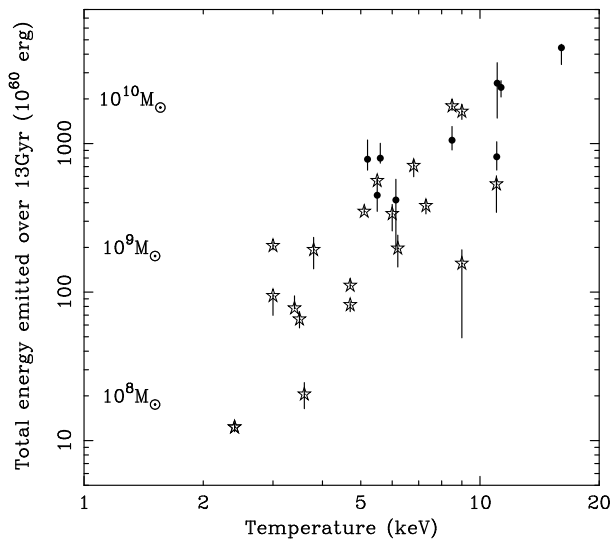}
\caption{Total energy required to stem cooling flows. The equivalent
accreted black hole masses, based on a radiative efficiency of 10 per
cent, are indicated. Symbols as in Fig.~1.}
\end{figure}

These constraints are shown in Fig.~2. $\Lambda_{\rm m}$ has been taken to
be $10^{-23}\ergcmcups$ and  $\kappa_0=5\times 10^{-7}\ergpcmps
\rm{K}^{-3.5}.$ We parametrize the gas fraction here as $f=0.1;$ the
fraction drops as stars form, or have formed at earlier stages.
Above the solid line, conduction can overwhelm radiative cooling and
below, cooling dominates. Dotted lines of constant mass from the
virial theorem, as well as a representative galaxy formation line
(dashed) where the radiative cooling time (from
Equation~\ref{eqn:tcool}) is $5\times 10^9\yr$, are shown. Conduction
can have a significant effect on the formation of galaxies with a
total mass exceeding a few times $10^{12}\Msun$. It is also relevant
to the work of Kauffmann et al.\ (1999) who assume that gas cooling in
potential wells with a circular velocity exceeding $350\kmps$ (which
corresponds to a temperature of about $10^7\K$) does not form visible
stars.

Conduction provides a simple explanation for the upper-mass cutoff in
galaxy formation. Something must be suppressing continued cooling in
massive haloes and conduction close to the Spitzer rate should become
effective in the required mass range. If the alternative is heating by
a central black hole, then it is not obvious why this becomes
effective in massive galaxies, where the atmosphere is most extensive,
and does not dominate in lower mass ones.

\section{Discussion}

The level of conductivity required to balance radiative cooling at the
cooling radius in most of the cores of 29 cooling flow clusters lies
within an order of magnitude of the Spitzer rate. This need not
indicate that conduction is operating but may merely be a coincidence,
given the age of the Universe and the universal gas fraction. The
detailed shape of the temperature profile and a comparison with
conduction models may be required to test whether conduction is a
significant factor in suppressing cooling flows. Essentially, Fig.~1
just shows that it is energetically feasible for conduction to play a
significant role in cluster cores.

The total energy required to suppress a flow can be considerable. We
plot this energy, determined from $L(<r_{\rm cool}) t$, in Fig.~3. In
the diagram $t$ is the age of the Universe, and may be reduced by a
factor of a few, but the energy is nevertheless large. Whatever the heat
source, it is to be among the major heat flows in the Universe. If
accretion onto the central black hole is responsible, with the energy
flowing out in jets (see e.g.\ Churazov et al.\ 2002, and references
therein), then assuming an efficiency of $0.1$ of the accreted rest
mass, the accumulated black hole masses lie between about $10^9$
and $10^{10}\Msun$, varying inversely with the assumed efficiency.
This is often taken as the radiative efficiency of
accretion, but it is likely to be significantly smaller for the total jet
power dissipated as heat into the surrounding gas. This may more than
counterbalance the likely factor of a few overestimate for the cluster
age (say 5~Gyr rather than the 13~Gyr assumed). We therefore consider
that the masses indicated are lower limits for this process.

Determining the extent to which conduction operates in cluster cores
from observation will be difficult. Cold fronts indicate that that it
is suppressed at least locally in some situations. The reasonable
agreement over a range of radii which we found in a study of the
$\kappa_{\rm eff}$ profile in two clusters (Voigt et al.\ 2002) is good
supporting evidence for conduction operating in general. Work is in
progress to study more clusters. Of great importance is the magnetic
field structure and its connectedness. If the intracluster medium
consists of many separate small magnetic structures then large scale
conduction would be inhibited except where reconnection allows
structures to merge. Faraday rotation studies indicate in some objects
that the fields have coherence scales of a few kpc (Carilli \& Taylor
2002 and references therein). The temperature profiles of clusters so
far appear to be smooth on the scales studied, although projection
effects are important, and observational exposures generally not long
enough to probe subtle variations. In some objects (e.g.\ the Perseus
cluster, Fabian et al.\ 2000; and the Centaurus cluster, Sanders \&
Fabian 2002) spiral-like X-ray structures are seen which could reflect
changes in conductivity associated with magnetic field structure.

\section*{Acknowledgments}
LMV and RGM acknowledge support from PPARC. ACF thanks the Royal Society
for support.


\begin{thebibliography}{}

\bibitem []{} Allen S.~W., 1995, MNRAS, 276, 947
\bibitem []{} Allen S.~W., 2000, MNRAS, 315, 269
\bibitem []{} Allen S.~W., Schmidt R.~W., Fabian A.~C., 2001, MNRAS,
  328, L37
\bibitem []{} B\"ohringer H., Matsushita K., Churazov E., Ikebe Y., 
  Chen Y., 2002, A\&A, 382, 804 
\bibitem []{} Br\"uggen M., Kaiser C.~R., 2001, MNRAS, 325, 676
\bibitem []{} Bregman J.~N., David L.~P., 1988, ApJ, 326, 639
\bibitem []{} Cardiel N., Gorgas J., Aragon-Salamanca A., 1998, MNRAS,
  298, 977
\bibitem []{} Carilli C.~L., Taylor G.~B., 2002, ARAA, in press (astro-ph/0110655)
\bibitem []{} Churazov E., Sunyaev R., Forman W., Böhringer H., 2002,
  MNRAS, 332, 729
\bibitem []{} Crawford C.~S., Allen S.~W., Ebeling H., Edge A.~C.,
  Fabian A.~C., 1999, MNRAS, 306, 857
\bibitem []{} De Grandi S., Molendi S., 2001, ApJ, 551, 153 
\bibitem []{} Ettori S., Fabian A.~C., 2000, MNRAS, 317, L57
\bibitem []{} Ezawa H., Fukazawa Y., Makishima K., Ohashi T., Takahara F.,
  Xu H., Yamasaki N.~Y., 1997, ApJ, 490, 33
\bibitem []{} Fabian A.~C., Pringle J.~E., 1977, MNRAS, 181, 5P
\bibitem []{} Fabian A.~C., Canizares C.~R., B\"ohringer H., 1994, ApJ, 425, 40
\bibitem []{} Fabian A.~C. et al., 2000, MNRAS, 318, L65
\bibitem []{} Fabian A.~C., Mushotzky R.~F., Nulsen P.~E~J., Peterson
  J.~R., 2001, MNRAS, 321, L20 
\bibitem []{} Fukazawa Y. et al., 1994, PASJ, 46, L55
\bibitem []{} Gilfanov M.~R., Sunyaev R.~A., 1984, SvAL, 10, 137
\bibitem []{} Gorgas J., Pedraz S., Guzman R., Cardiel N., Gonzalez J.~J.,
  1997, ApJ, 481, L19
\bibitem []{} Gruzinov A., 2002, astro-ph/0203031
\bibitem []{} Johnstone R.~M., Fabian A.~C., Nulsen P.~E.~J., 1987, MNRAS,
  224, 75 
\bibitem []{} Johnstone R.~M.,  Allen S.W., Fabian A.~C., Sanders J.S.,
  2002, MNRAS, in press (astro-ph/0202071)
\bibitem []{} Kauffmann G.,  Colberg J.M., Diaferio A., White  S.D.M.
  1999, MNRAS, 303,188
\bibitem []{} Loeb A., 2002, NewA, 7, L279
\bibitem []{} Lynden-Bell D., 1967, Obs, 87, L163
\bibitem []{} Malyshkin L., 2001, ApJ, 554, 561
\bibitem []{} Markevitch M. et al., 2000, ApJ, 541, 542
\bibitem []{} McNamara B.~R. et al., 2001, ApJ, 562, 149    
\bibitem []{} Narayan R., Medvedev M.~V., 2001, ApJ, 562, L129
\bibitem []{} Navarro J.~F., Frenk C.~S., White S.~D.~M., 1997, ApJ, 490, 493
\bibitem []{} Nulsen P.~E.~J., David L.~P., McNamara B.~R., Jones C.,
  Forman W.~R., Wise M., 2002, ApJ, 568, 163
\bibitem []{} Peres C.~B., Fabian A.~C., Edge A.~C., Allen S.~W.,
  Johnstone R.~M., White D.~A., 1998, MNRAS, 298, 416
\bibitem []{} Peterson J.~R. et al., 2001, A\&A, 365, L104
\bibitem []{} Qin B., Wu X., 2000, ApJ, 529, L1
\bibitem []{} Reynolds C.~S., Heinz S., Begelman M.~C., 2002, MNRAS, 332, 271
\bibitem []{} Sanders J.~S., Fabian A.~C., 2002, MNRAS, 331, 273 
\bibitem []{} Soker N., Sarazin C.~L., 1990, ApJ, 348, 73
\bibitem []{} Spitzer L., 1962, Physics of Fully Ionized Gases, New
  York: Wiley-Interscience
\bibitem []{} Tamura T. et al., 2001, A\&A, 365, L87
\bibitem []{} Vikhlinin A., Markevitch M., Murray S.~S., 2001, ApJ, 551, 160
\bibitem []{} Voigt L.~M., Schmidt R.~W., Fabian A.~C., Allen S.~W.,
  Johnstone R.~M., 2002, MNRAS, in press (astro-ph/0203312)
\bibitem []{} White S.~D.~M., Frenk C.~S., 1991, ApJ, 379, 52
\end{thebibliography}
\end{document}